\documentclass[10pt, conference]{IEEEtran}
\usepackage{cite}
\usepackage{amsmath,amssymb,amsfonts}
\usepackage{amsthm}
\usepackage[ruled]{algorithm2e}
\usepackage{algpseudocode}
\usepackage{graphicx}
\usepackage{textcomp}
\usepackage{xcolor}
\usepackage{subcaption}
\usepackage{multirow}
\usepackage{booktabs}
\usepackage{makecell}
\usepackage{bbm}
\usepackage{threeparttable}

\newcommand{\linebreakand}{%
  \end{@IEEEauthorhalign}
  \hfill\mbox{}\par
  \mbox{}\hfill\begin{@IEEEauthorhalign}
}

\begin{document}
\title{Leveraging Reinforcement Learning in Red Teaming for Advanced Ransomware Attack Simulations}

\author{
        Cheng Wang$^{a}$$^{*}$,
        Christopher Redino$^{a}$,
        Ryan Clark$^{a}$,
        Abdul Rahman$^{a}$,
        Sal Aguinaga$^{a}$
        Sathvik Murli$^{a}$,\\
        Dhruv Nandakumar$^{a}$,
        Roland Rao$^{a}$,
        Lanxiao Huang$^{b}$,
        Daniel Radke$^{a}$,
        Edward Bowen$^{a}$\\
        \small $^{a}$Deloitte \& Touche LLP \\
        \small $^{b}$National Security Institute, Virginia Tech \\
        \small $^{*}$Corresponding author: chengwang@deloitte.com \\
}


\maketitle

\begin{abstract}
Ransomware presents a significant and increasing threat to individuals and organizations by encrypting their systems and not releasing them until a large fee has been extracted. To bolster preparedness against potential attacks, organizations commonly conduct red teaming exercises, which involve simulated attacks to assess existing security measures. This paper proposes a novel approach utilizing reinforcement learning (RL) to simulate ransomware attacks. By training an RL agent in a simulated environment mirroring real-world networks, effective attack strategies can be learned quickly, significantly streamlining traditional, manual penetration testing processes. The attack pathways revealed by the RL agent can provide valuable insights to the defense team, helping them identify network weak points and develop more resilient defensive measures. Experimental results on a 152-host example network confirm the effectiveness of the proposed approach, demonstrating the RL agent's capability to discover and orchestrate attacks on high-value targets while evading honeyfiles (decoy files strategically placed to detect unauthorized access).
\end{abstract}


\section{Introduction}
As a form of malicious software, ransomware aims to encrypt files or lock computer systems until a ransom is paid (typically in cryptocurrency) to the attacker. It is a form of cyber extortion where the attacker demands payment in exchange for restoring access to the encrypted files or unlocking the affected system. Ransomware attacks are increasingly prevalent and sophisticated, and have targeted individuals, businesses, and governmental institutions \cite{razaulla2023age, benmalek2024ransomware}, with notable incidents in recent years including the WannaCry outbreak in 2017 affecting thousands of systems worldwide \cite{chen2017automated}.

Mitigating ransomware attacks can be accomplished through regularly conducting \emph{red teaming} exercises \cite{mansfield2018best} aimed at simulating such attacks to enhance preparedness. These exercises serve to test the efficacy of existing security measures and infrastructure. Organizations can develop and implement targeted security measures by gaining insight into potential vulnerabilities within their systems, empowering them to better safeguard against ransomware threats and mitigate associated risks.

One approach to achieving this involves constructing a simulated environment, often referred to as a ``digital twin", that mirrors the real network infrastructure. Instead of enumerating all potential attack pathways or manually selecting them based on deep domain expertise, an RL agent can be trained to \emph{automatically} identify effective attack strategies by directly interacting with the simulated environment. Leveraging RL not only lessens the burden of manual testing, thereby enhancing the efficiency and scalability of penetration testing, but also enables adaptive learning and continual refinement of defense strategies. The findings generated by the RL agent can provide valuable insights that enable the defense (blue) team to refine existing strategies or develop new ones. The new environment can then be leveraged to train new agents, further testing the robustness of the defense measures. Through this iterative learning and improvement process, weaknesses in the defense mechanisms can be continuously identified and addressed. Consequently, this iterative approach can help foster a proactive defense posture, tested against objective-driven (as opposed to rule-based) dynamic attacking strategies, thereby enhancing resilience against ransomware threats. 

To our knowledge, this study represents the first exploration of using RL to simulate ransomware attacks at the network level. We introduce a novel ransomware attack simulation model designed to replicate real-world scenarios where attackers seek to encrypt sensitive hosts within a limited time frame. The attacker's strategic dilemma is central to our model: exploring the network to uncover valuable targets or immediately encrypt known hosts. The latter risks triggering prompt defense responses and prematurely ending the attacker's campaign. Additionally, the attacker must navigate the network while avoiding encrypting honeyfiles\cite{yuill2004honeyfiles}. This complex decision-making process underscores the nontrivial nature of coordinating attacks on high-value assets. A detailed RL formulation for this attack simulation is then presented. Experiment results on a 152-host medium-sized network demonstrate the effectiveness and efficiency of the attack strategies learned by the RL agents with different risk profiles. In addition, we show how, based on the initial results, honeyfiles can be strategically added to improve security and how, by retraining the RL agent in the new environment, additional insights are revealed, which may help the blue team continue to devise and implement more robust defense measures.

This paper is organized as follows: Section \ref{sec:related} reviews related work on ransomware and RL-based penetration testing. Section \ref{sec:preliminaries} provides the necessary background on RL. The attack simulation and its RL model are presented in Section \ref{sec:main}. Section \ref{sec:experiments} details the experimentation setup and results. Finally, section \ref{sec:conclusion} concludes by discussing the limitations of the current work and suggesting avenues for future research.

\section{Related Work} \label{sec:related}
Increasing interest in the application of machine learning (ML) to detection and classification of ransomware \cite{oz2022survey, razaulla2023age} has led to multitude of classifiers being proposed, with notable algorithms or architectures including support vector machine \cite{takeuchi2018detecting, ahmed2021peeler}, random forest \cite{almousa2021api, khammas2020ransomware}, convolutional neural network \cite{basnet2021ransomware}, long short-term Memory \cite{molina2021ransomware}, and self-attention \cite{zhang2020ransomware, roy2021deepran}. These models are trained on various features, such as application programming interface (API) calls, network traffic, and file entropy, which are often derived through dynamic analysis where a sample of malicious code is executed in a controlled environment \cite{damodaran2017comparison}. These ML based approaches have demonstrated high accuracy along with a low false-positive rate.

However, the exploration of utilizing RL for ransomware simulation is still in its nascent stages. In \cite{adamov2020reinforcement}, an RL-based ransomware simulator is proposed to evade a rule-based detection system on a single host machine. The RL agent aims to encrypt as many files as possible while keeping the number of files with a second extension, high entropy, or similar modification time below a given threshold to avoid detection. In their experiments, there are only 11 distinct states in total, which correspond to the number of encrypted files (i.e., 0, 1, ..., 10). The action space is also limited, consisting of just 16 possible actions, based on a combination of three factors: the number of files to encrypt (such as 1, 2, 5, or 10), the decision to add an extension to encrypted files, and the choice to encode files with Base64 to reduce their entropy. The results showcased the potential of using RL to identify attack strategies that can bypass an existing rule-based detection system on individual hosts.

Other notable applications of RL in red teaming exercises include crown-jewel  (CJ) analysis \cite{gangupantulu2021crown}, exfiltration (exfil) path discovery \cite{cody2022discovering, rishu2023enhancing}, surveillance detection routes (SDR) \cite{huang2022exposing}, and command and control (C2)\cite{10115173}. These studies integrate cyber terrain \cite{conti_raymond_2018, gangupantulu2021using} into the RL environment through meticulous but manual reward engineering and adjustment of transition probabilities. In these works, the underlying networks are derived directly from scanning results of enterprise networks, typically comprising hundreds to thousands of hosts. Experiment results have shown the effectiveness and efficiency of using RL to automate the discovery process of attack paths in various penetration testing tasks.

\section{Preliminaries on Reinforcement Learning} \label{sec:preliminaries}
In RL, an agent learns to optimize its behaviors by interacting with the environment \cite{sutton2018reinforcement}. The environment is defined as a Markov decision process (MDP): $(\mathcal{S}, \mathcal{A}, \mathcal{P}, r, \gamma)$, where $\mathcal{S}$ is the state space, $\mathcal{A}$ is the action space, $\mathcal{P}: \mathcal{S} \times \mathcal{A} \times \mathcal{S}\rightarrow [0, 1]$ is the transition probability function, $r: \mathcal{S} \times \mathcal{A} \times \mathcal{S} \rightarrow \mathbb{R}$ is the reward function and $\gamma \in (0,1]$ is the discount factor, which determines the present value of future rewards.
At each step $t$, the agent observes a state $s_t$ and selects an action $a_t$ according to its policy $\pi: \mathcal{S} \times \mathcal{A} \rightarrow [0, 1]$, which is a probabilistic distribution over each action. It then transitions to the next state $s_{t+1}$ according to $\mathcal{P}$ and receives a reward $r_t=r(s_t, a_t, s_{t+1})$. The goal of the agent is to learn a policy that maximizes the expected cumulative rewards $G_t\equiv\sum_{k=0}^\infty \gamma^k r_{t+k}$, which is also referred to as \emph{return}. 

One approach to learning a policy involves optimizing a performance measure $J(\theta)$, such as the expected return, where $\theta$ is the parameters of the policy. According to the Policy Gradient Theorem \cite{sutton1999policy}, the gradient of the policy can be estimated as 
\begin{align}
    \nabla_\theta J(\theta)  \approx \mathbb{E}\Big[\sum_{t=0}^{T}\nabla_{\theta}\log_\theta \pi(a_t|s_t)A^{\pi_\theta}(s_t, a_t)\Big],
\end{align}
where $A^{\pi_\theta}(s_t, a_t)=Q^{\pi_\theta}(s_t, a_t) - V^{\pi_\theta}(s_t)$ is the \emph{advantage} function. Here $V^{\pi_\theta}(s_t)$ is the expected return from state $s_t$ and $Q^{\pi_\theta}(s_t, a_t)$  represents the state-action value, i.e., the expected return of taking action $a_t$ in state $s_t$ and thereafter following policy $\pi_\theta$. The advantage function quantifies the relative benefit of choosing a specific action $a_t$ in a given state $s_t$, compared to the expected value of the average action under the current policy. 

To improve training stability and avoid large updates between steps, the Proximal Policy Optimization (PPO) algorithm \cite{schulman2017proximal} instead uses a clipped surrogate objective function:
\begin{align}
    \mathcal{L}(\theta) = \mathbb{E} \Big[ \min\big(\rho_t(\theta) A_t, \mathrm{clip}\big( \rho_t(\theta), 1-\epsilon, 1+ \epsilon \big) A_t\big)\Big], \label{eq:ppo_obj}
\end{align}
where $\rho_t(\theta) = \pi_\theta(a_t|s_t) / \pi_{\theta_{\mathrm{old}}} (a_t|s_t)$ is the probability ratio of the new policy over the old policy. The advantage function $A_t$ in \eqref{eq:ppo_obj} can be estimated using the generalized advantage estimation (GAE) \cite{schulman2015high}, truncated after $T$ steps:
\begin{align}
    & \hat{A}_t = \delta_t + (\gamma\lambda) \delta_{t+1}+\cdots + (\gamma\lambda)^{T-t+1}\delta_{T-1},\\
    & \mathrm{where\;} \delta_t = r_t + \gamma V(s_{t+1}) - V(s_t).
\end{align}
Finally, as a means to encourage sufficient exploration, an entropy bonus $\beta H(\pi_\theta)$ is often added to the objective function \eqref{eq:ppo_obj}, where $\beta$ is a hyperparameter. 
The PPO algorithm is shown in Algorithm \ref{alg:ppo} for reference.

\begin{algorithm}[t]
    \caption{Proximal Policy Optimization (PPO)}\label{alg:ppo}
    \SetKwInOut{Initialize}{Initialize}
    \Initialize{actor network $\pi_\theta$  and critic network $V_w$}
    \For{$\mathrm{iteration = 1,2,...}$}{
        Set $\theta_{\mathrm{old}}=\theta$ \\
        Run $\theta_{\mathrm{old}}$ for T steps \\
        Compute advantage estimates $\hat{A}_t$ and targets $\hat{V}_t$\\
        \For{$\mathrm{epoch = 1,2,...,K}$}{
             Compute surrogate objective $\mathcal{L}^{\mathrm{act}}(\theta)$ \\
             Compute entropy $H(\theta)$\\
             Compute mean-squared-error $\mathcal{L}^{\mathrm{crt}}(w)$\\
             Update actor $\theta \leftarrow \theta + \alpha_\theta\nabla\big[{\mathcal{L}^{\mathrm{act}}(\theta)} + \beta H(\theta)\big]$\\
             Update critic $w \leftarrow w - \alpha_w\nabla{\mathcal{L}^{\mathrm{crt}}(w)}$
        }
    }
\end{algorithm}
\
\section{Reinforcement Learning Model for Ransomware Attack Simulation} \label{sec:main}
\subsection{Attack Simulation Overview}
We consider a scenario where an attacker, having gained an initial foothold on a host in a network, seeks to identify, compromise, and encrypt high-value targets in a swift and stealthy manner for maximum impact and ransom leverage. 
In this red teaming exercise, we aim to emulate adversary strategies tailored to evade deployed defensive measures and to identify the optimal time and placement for delivering offensive capabilities.

It is assumed that different hosts in a given network have varying levels of importance, each characterized by a numerical value. Upon getting initial access to a network, attackers often refrain from immediate aggressive actions to evade alerting by defensive infrastructure. Instead, adversaries initiate careful and quiet exploration while moving laterally to identify the most valuable assets. By avoiding early detection, the chances of accessing high-value targets and inflicting more significant damage are quite high, thereby maximizing effectiveness and payoff.

The following defensive mechanisms are assumed to be in place based on information technology (IT) leading practices \cite{hochstein2005-itil}. First, it is assumed that each system is regularly updated and patched. This causes attackers to  access to compromised hosts after a period of time if no further action is taken. Consequently, they have only a limited time window to perform follow-up actions such as scanning or encrypting the underlying host before losing their initial access. Second, once a device is encrypted and the ransom note is posted, there is a delay (e.g., one hour) before the defense team isolates it from the rest of the network. During this time, the attacker may continue to move laterally, discovering and compromising adjacent systems. Thirdly, honeyfiles are strategically deployed on selected hosts. Attempts to encrypt these files will immediately trigger an alert, leading to the swift isolation of the affected host. Finally, when the total number of isolated hosts exceeds a predefined threshold, in scope systems will be taken offline and examined. This measure prevents further damage from occuring and effectively terminates the current attack campaign.

In addition to the traditional time steps in MDPs, a simulated wall clock tracks the progress of attacks and responses. Different actions require varying amounts of time to execute, and certain defense measures, as previously described, automatically activate upon meeting a predefined criterion.


\subsection{States}
The state of the RL environment consists of a number of features from each host in the network, such as its address, operating system (OS), running services and processes, infection and encryption status, and so on (see Table \ref{tab:states} for a list). It is important to note whether a host has honey files (i.e., \texttt{HAS\_Honeyfiles}) is not directly observable to the RL agent. However, the agent can potentially discover honey files with a certain probability by taking a \texttt{File\_Scan} action. If honey files are detected on a host, the corresponding feature \texttt{Found\_Honeyfiles}, observable to the agent, will be set to \texttt{True}.
\begin{table}[t]
    \centering
    \caption{Overview of the state space.}
    \begin{tabular}{l|l} \hline
       \textbf{Host Feature}  & \textbf{Description} \\ \hline
       Subnet\_ID  & One-hot encoding of the host's subnet.\\
       Local\_ID  & One-hot encoding of the host's local address.   \\
       Discovered & Whether the host has been discovered.\\
       OS & Whether an OS is running on the host.\\
       Services &  Whether a service is running on the host.\\
       Processes & Whether a process is running on the host.\\
       Has\_Honeyfiles & Whether the host has honeyfiles. \\
       Found\_Honeyfiles & Whether honeyfiles has been found on the host.\\
       Exploited & Whether the host has been compromised.\\
       Access & Access level gained on the host.\\
       Scanned & Whether the host's file system has been scanned.\\
       Encrypted & Whether the host has been encrypted.\\
       Isolated & Whether the host has been isolated by defenders.\\
       Value & Value of encrypting the host.\\
       Infection\_Time & Time elapsed since the host was compromised.\\
       Encryption\_Time& Time elapsed since the host was encrypted. \\
       \hline
    \end{tabular}
    \label{tab:states}
\end{table}

\subsection{Actions}
\begin{table}[b]
    \centering
    \caption{Overview of the action space.}
    \begin{tabular}{l|l} \hline
       \textbf{Action}  &  \textbf{Description} \\ \hline
       Subnet\_scan  & Discover nearby hosts and their configurations.\\ 
       Exploit & Exploit a vulnerability to establish a foothold.\\
       File\_scan & Scan file system to identify potential honeyfiles.\\
       Encrypt & Encrypt file system to restrict access by original users.\\
       \hline
    \end{tabular}
    \label{tab:actions}
\end{table}
There are four types of actions in the action space (Table \ref{tab:actions}). 
The first type, \texttt{Subnet\_scan}, is used to discover neighboring hosts and find out their OS, running services and processes. Access to the underlying host must be gained to perform a subnet scan. A successful scan will reveal each of the hosts within the same subnet, as well as some hosts from adjacent subnets, subject to firewall rules. 

Next, a suitable \texttt{Exploit} action must be selected and executed to gain control of a discovered host. Each exploit targets a specific OS, service, or process, and is associated with a corresponding vulnerability in the Common Vulnerabilities and Exposures database. 

After establishing a foothold, the agent may conduct a thorough scan of the host's file system by initiating a \texttt{File\_scan} action, enabling the agent to assess the host's significance and identify potential honeyfiles. 
This may be done through metadata analysis, without directly accessing or opening the files themselves. 
If honeyfiles are present on a system, the agent can detect them with a certain probability and then set the host's attribute \texttt{Found\_Honeyfiles} to \texttt{True}.

Finally, the agent can initiate the encryption process by executing the \texttt{Encrypt} action. When honeyfiles are modified, alerts will be triggered and the underlying host will be isolated to prevent further unauthorized activities or network infiltration. 
However, if honeyfiles have been identified on a system, the attacker agent will skip those files and encrypt the remaining files without triggering an alert.
As a means to expedite the attack campaign, the agent may choose to bypass the file scanning step for some hosts and proceed directly to the encryption phase, although this may lead to undesirable consequences, such as accessing decoy files or encrypting hosts of lesser significance.

\subsection{Rewards}
The reward function encompasses a positive component $r^+$, rewarding successful outcomes in discovery, exploitation, and encryption, coupled with a negative component, $\rho r^-$, accounting for the associated costs of traversing cyber terrain, where $\rho$ symbolizes the level of the risk aversion:
\begin{align}
    r(s,a,s') = r(s,a,s')^+ - \rho r^-(s,a,s'). \label{eq:reward}
\end{align}

It is worth mentioning that the penalty term depends not only on the type of the action but also on its target host. As devices within a network typically serve different roles and functionalities, the defensive measures around them usually vary. For instance, critical assets such as databases are usually more closely monitored than other systems. Although the exact defense measures in place may not be known, experienced hackers or penetration testers can often deduce the level of security surrounding a target based on the services it is running. By varying the penalties for taking actions towards different targets, the reward function \eqref{eq:reward} hence incorporates the heterogeneity in cyber terrains as introduced in \cite{gangupantulu2021using}.

\section{Experimentation and Results} \label{sec:experiments}

\subsection{Network Configuration}

\begin{figure}[t]
    \centering
    \includegraphics[width=0.48\textwidth]{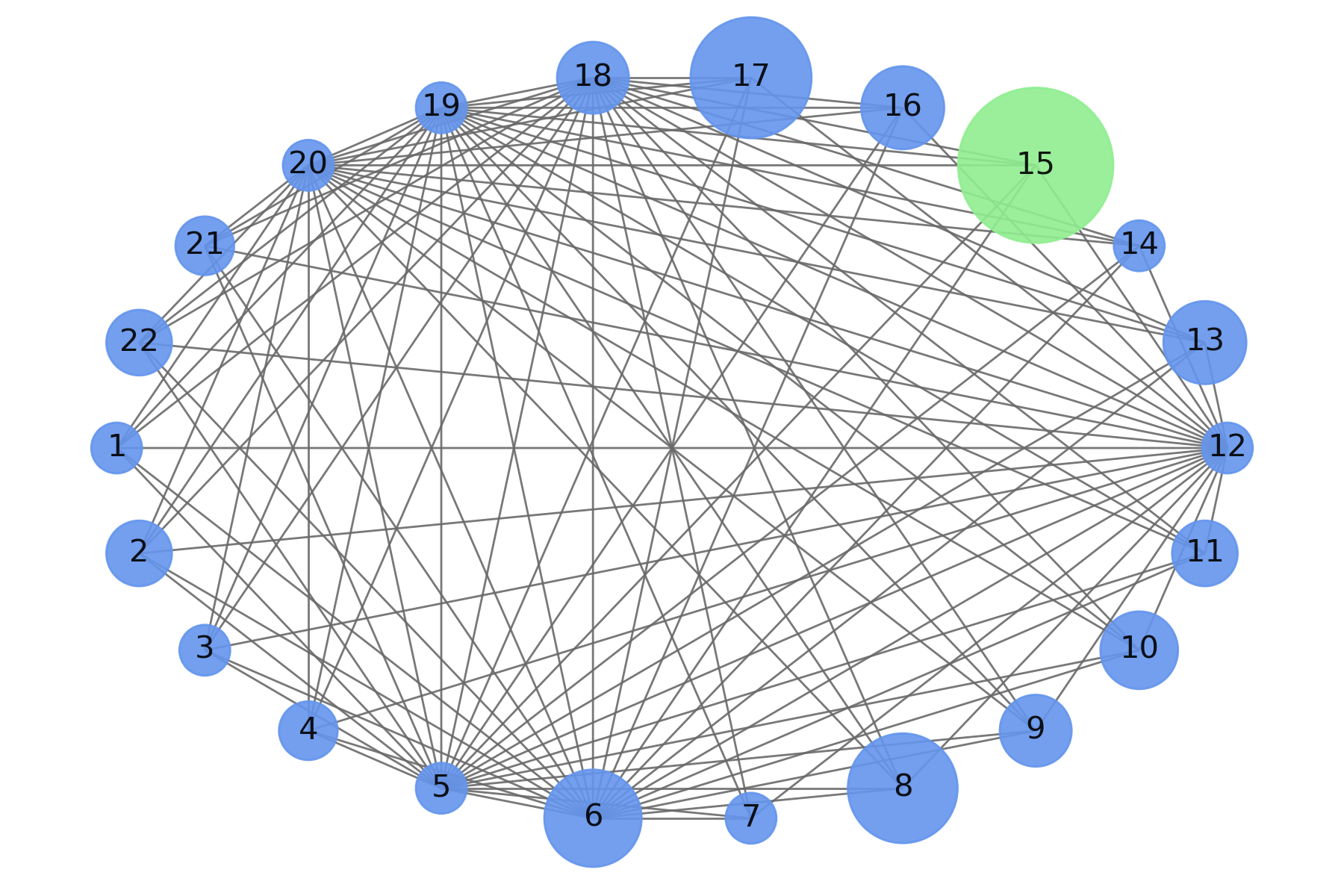}
    \caption{Experiment network topology overview: Nodes depict \emph{subnets}, with sizes proportional to the number of hosts within. Only subnet 15 (green node) is public.}
    \label{fig:network}
\end{figure}

The experimental network is designed by domain professionals to emulate a mid-size enterprise network, intentionally incorporating certain weaknesses. 
The entire network (Fig. \ref{fig:network}) comprises 152 hosts distributed across 22 subnets, each representing distinct functionalities or areas of operation. In particular, subnet 15 serves as the demilitarized zone (DMZ), representing the network's sole public-facing subnet. It is assumed that the attacker has gained an initial foothold on host (15, 12), where 12 is the host's local ID in subnet 15. This host provides multiple services, including web hosting, mail, and file transfer.

Out of the 152 hosts across the whole network, 14 have been designated as sensitive hosts, and 23 have been placed with honeyfiles. It's important to note that not all hosts are reachable or exploitable by the RL agent. It is the agent's responsibility to identify vulnerable high-value targets and coordinate its attacks accordingly.

\subsection{Training setup}
The rewards for successfully discovering, exploiting, and encrypting a host are 10, 10 and 50, respectively. In addition, upon successfully encrypting a host, the agent receives a bonus reward equivalent to the predetermined value assigned to the host. This value ranges from 1000 for sensitive hosts to 0 for non-critical hosts. The penalty for an action depends on its type and the services running on its target host, ranging from 1 to 6. An episode ends when more than 3 hosts are flagged or isolated. A host is immediately flagged if its honeyfiles are accessed. Otherwise, if the host is encrypted, it will be isolated from the rest of the network within one hour, assuming there is a delay in the response from defense teams. The wall-clock times for the actions subnet scan, exploit, file scan, and encrypt are set to 30, 10, 60, and 300 seconds, respectively.

Three RL models are trained using the PPO algorithm \cite{schulman2017proximal}, as listed in Algorithm \ref{alg:ppo}, with varying risk aversion factors: low ($\rho=1$), medium ($\rho=5$), and high ($\rho=20$).
Both actor (policy) and critic (value) networks consist of 2-layer multi-perceptron neural networks with 200 neurons in each layer. Table \ref{tab:hyperparams} lists the main hyperparameters used for training. 

\begin{table}[t]
    \centering
    \caption{PPO hyperparameters.}
    \begin{tabular}{l|l} \hline
        \textbf{Hyperparameter} & \textbf{Value} \\ \hline
        Horizon (T) & 4096 \\ 
        Minibatch size & 64 \\
        Epochs & 10 \\
        Clipping ratio ($\epsilon$) & 0.1 \\
        Entropy bonus ($\beta$) & 0.005 \\ 
        Discount factor ($\gamma$) & 0.999 \\
        Advantage discount factor  ($\lambda$) & 0.95 \\
        Learning rate (critic) & $3\times10^{-4}$ \\ 
        Learning rate (actor) & $1\times10^{-5}$\\
        \hline
    \end{tabular}
    \label{tab:hyperparams}
\end{table}

\subsection{Results}
\begin{figure}[t]
    \centering
    \includegraphics[width=0.48\textwidth]{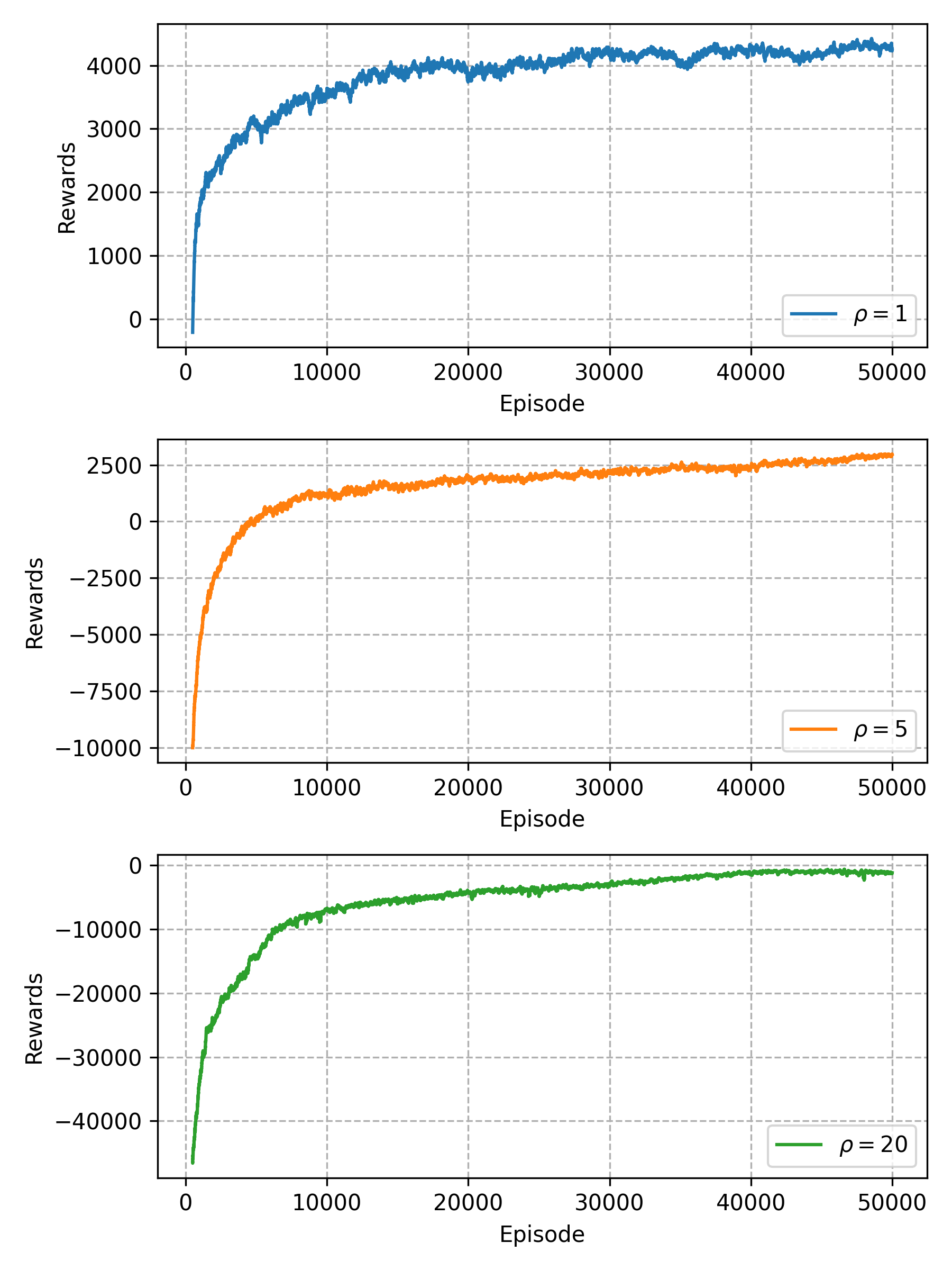}
    \caption{Episode rewards under different risk aversion factors: $\rho=1$ (top), $\rho=5$ (middle), and $\rho=20$ (bottom).}
    \label{fig:rewards}
\end{figure}

\begin{figure}[t]
    \centering
    \includegraphics[width=0.48\textwidth]{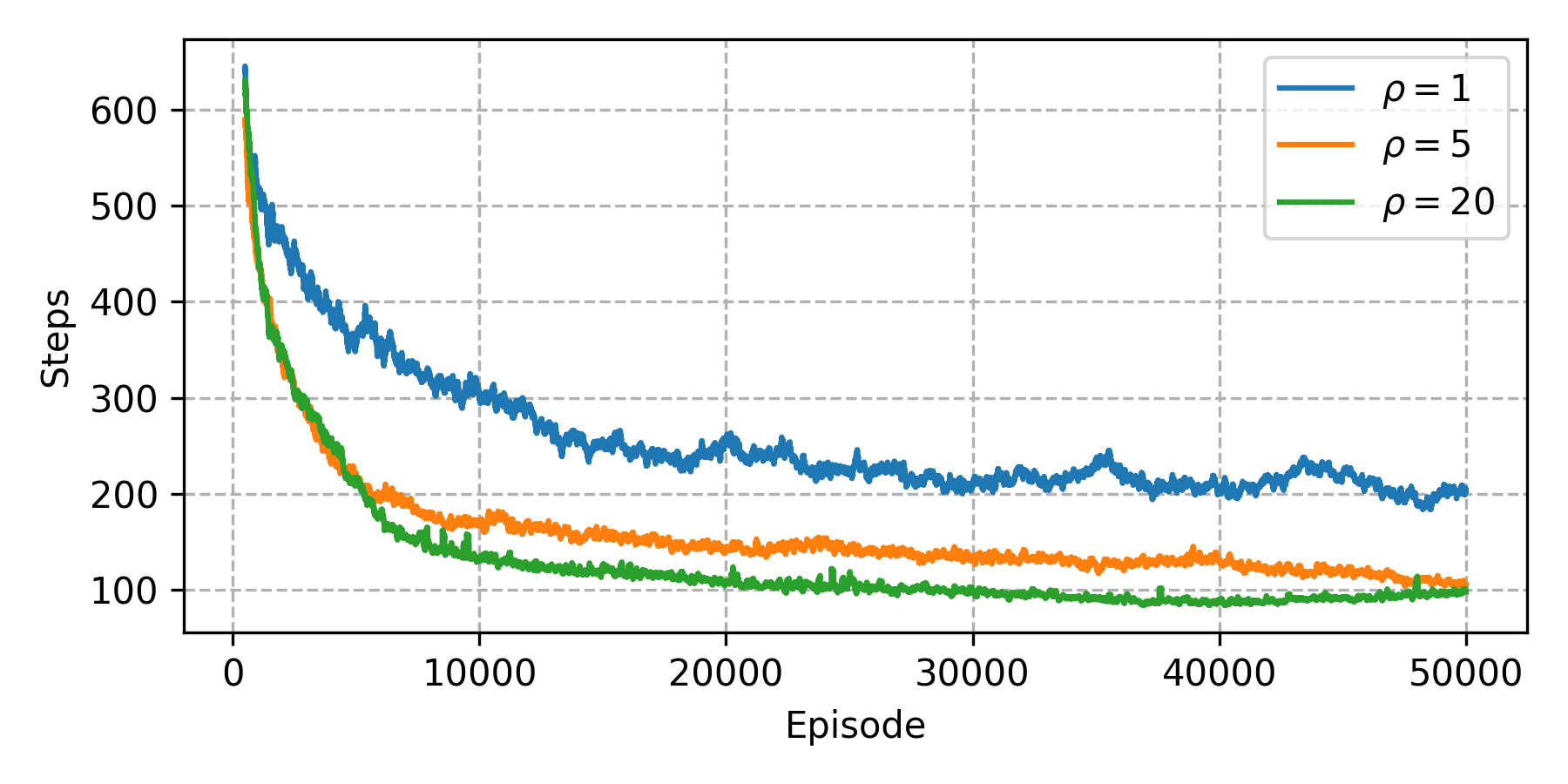}
    \caption{Episode lengths under different risk aversion factors.}
    \label{fig:steps}
\end{figure}

The three models, each with different risk aversion factors, converged within 50,000 episodes, as evidenced by the gradual increase in episode rewards depicted in Fig. \ref{fig:rewards}. Furthermore, the number of steps in an episode decreases as training progresses, as illustrated in Fig. \ref{fig:steps}, indicating the improved efficiency of the RL agent's operations. It is worth noting that the agent with high-risk aversion takes fewer steps compared to the agent with low-risk aversion. This is due to actions incurring increased penalties, prompting the agent to avoid unnecessary or exploratory steps.

\begin{figure}[t]
    \centering
    \includegraphics[width=0.48\textwidth]{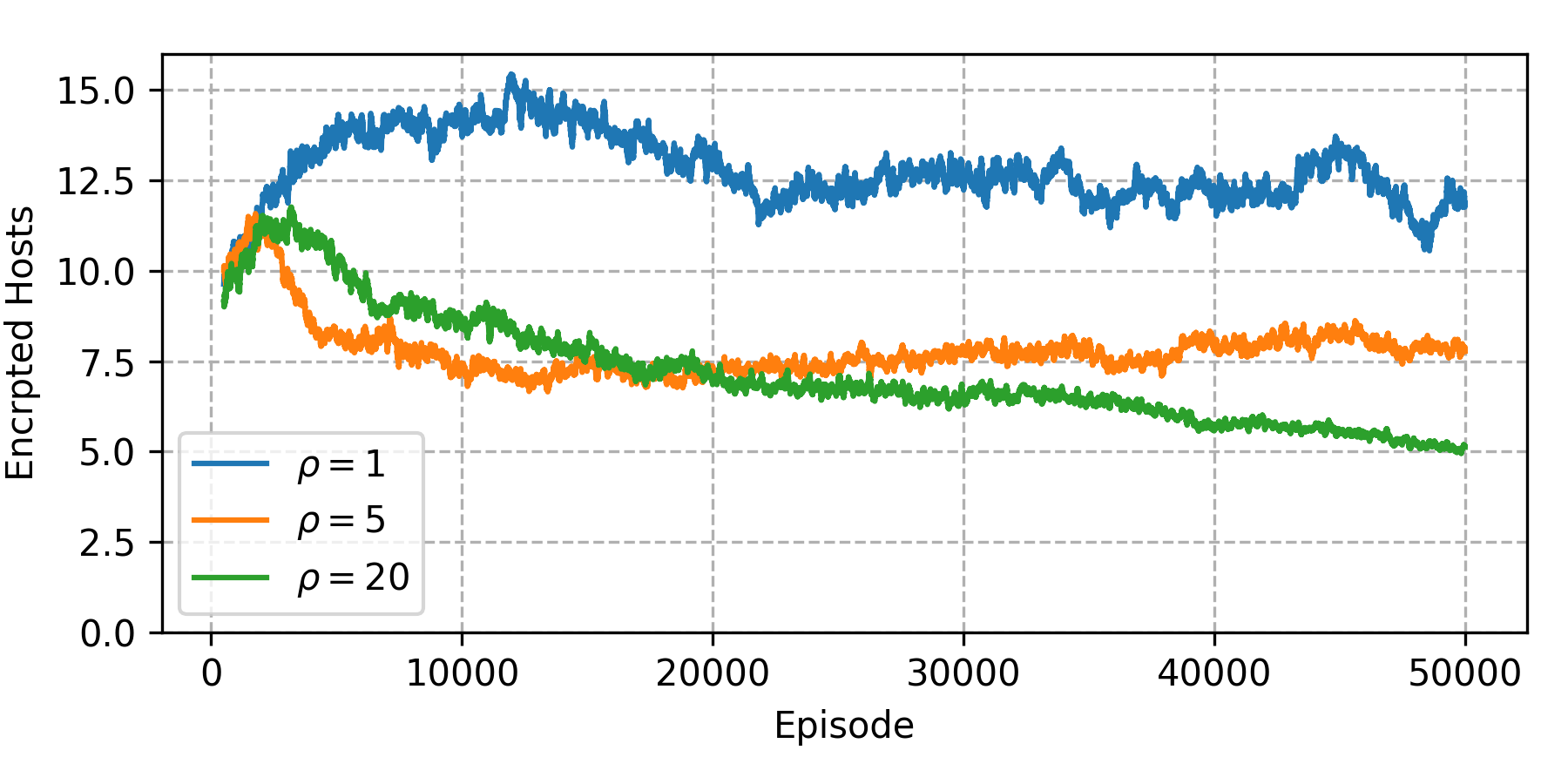}
    \caption{Number of encrypted hosts under different risk aversion factors.}
    \label{fig:encpt_hosts}
\end{figure}

Higher risk aversion also results in a lower number of encrypted hosts, as shown in Fig. \ref{fig:encpt_hosts}. With $\rho=20$, the agent encrypts just over 5 hosts as training approaches 50,000 episodes, whereas under $\rho=1$, the number of encrypted hosts more than doubles to over 12.

\begin{figure*}[t]
    \centering
    \includegraphics[width=0.96\textwidth]{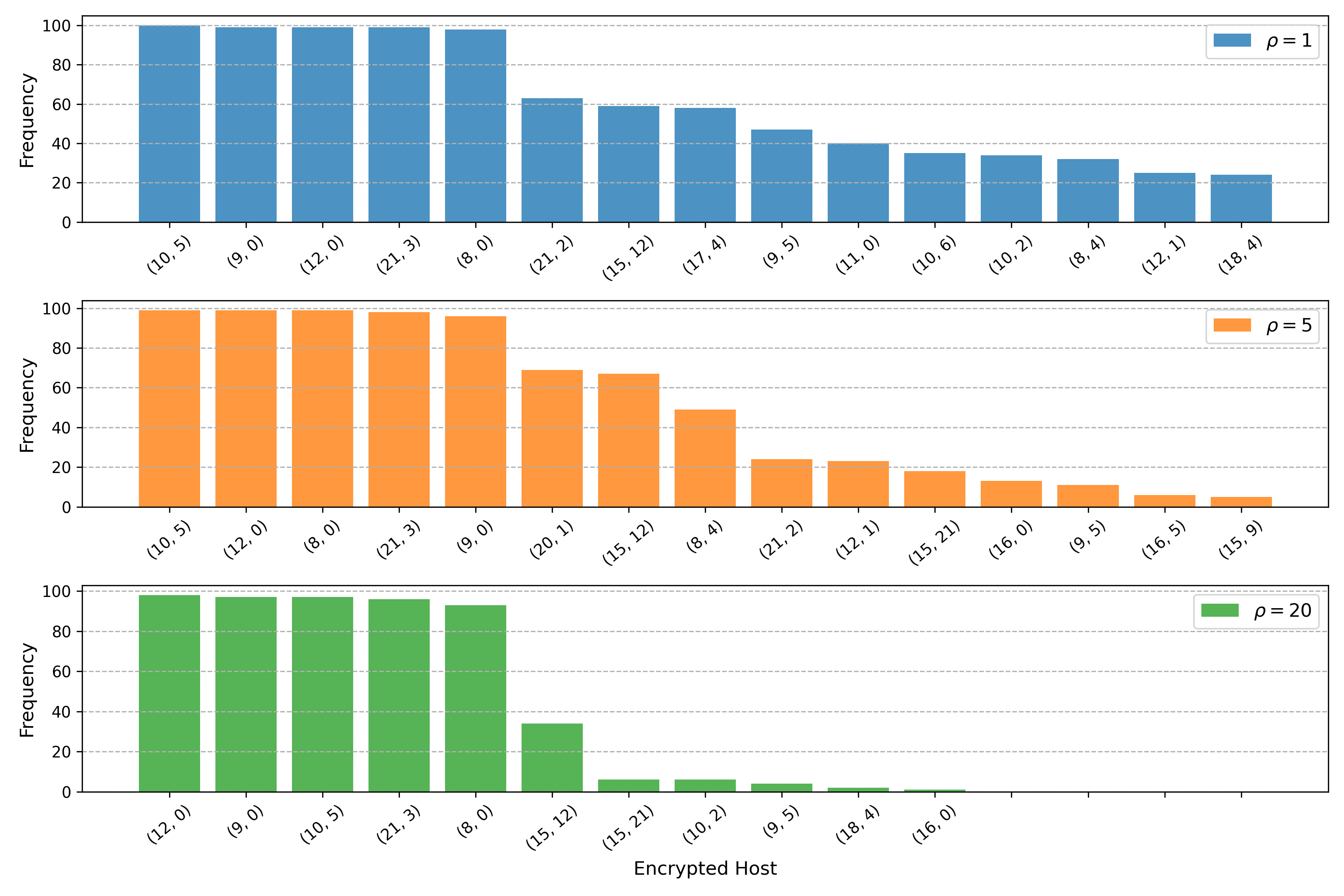}
    \caption{Top 15 most frequently encrypted hosts from 100 attack paths under different risk-aversion profiles: $\rho=1$ (top), $\rho=5$ (middle), $\rho=20$ (bottom).}
    \label{fig:encrypted_hosts_freq.png}
\end{figure*}

We generated 100 trajectories using the final model for each risk aversion factor to delve deeper into the hosts being encrypted under various risk profiles. Fig. \ref{fig:encrypted_hosts_freq.png} displays the most frequently encrypted hosts under each factor. Notably, five hosts consistently stand out, being encrypted more than 90 percent of the time, irrespective of the agent's risk preference. These hosts, namely (8, 0), (9, 0), (10, 5), (12, 0), and (21, 3), are sensitive hosts valued at 1000. Table \ref{tab:hosts_info} presents an overview of the key services operating on these hosts alongside their respective roles. As evident from the table, these hosts fulfill critical roles within the network, encompassing functions such as authentication, database management, file transfer, logging, network configuration, and printing services.

Another notable observation from Fig. \ref{fig:encrypted_hosts_freq.png} is that the agent sometimes encrypts host (15, 12), where it gains the initial access to the network, despite it being a non-sensitive host valued at 0 and yielding a reward of only 50 upon successful encryption.  However, as the penalty factor increases, this occurs less often due to the increased costs and diminishing incentives.

\begin{table*}[t]
    \centering
    \caption{Sensitive hosts overview: roles and key services.}
    \begin{threeparttable}
    \begin{tabular}{p{1cm}|p{7cm}|p{8cm}} \hline
       \textbf{Host}  & \textbf{Role} & \textbf{Key Services}\tnote{*}  \\ \hline
       (8, 0) &  Web services and secure access & HTTPS, Terminal server, OpenSSH\\
       (9, 0) & Network services and file sharing & NetBIOS-DGM, FTP\\ 
       (10, 5) & Database and logging services & SQL monitor, SYSLOG, Subroutine caller\\
       (12, 0) & Network management and file sharing & NetBIOS-SSN, SNMP, FTP \\
       (21, 3) & Network configuration and printing services & DHCP, Directory service, IPP \\
       \hline
    \end{tabular}
    \begin{tablenotes}\scriptsize
    \item[*] Acronyms of services refer to the following: 
    DHCP (Dynamic Host Configuration Protocol),
    FTP (File Transfer Protocol),
    HTTPS (Hypertext Transfer Protocol Secure),
    IPP (Internet Printing Protocol),
    NetBIOS-DGM (Network Basic Input/Output System Datagram service),
    NetBIOS-SSN (NetBIOS-Session Service),
    OpenSSH (Open Secure Shell),
    SNMP (Simple Network Management Protocol),
    SYSLOG (System Logging),
    \end{tablenotes}
    \end{threeparttable}
    \label{tab:hosts_info}
\end{table*}

\subsection{Impacts of Introducing New Honeyfiles}
By analyzing the behaviors and outcomes of the RL agents, the blue team can devise and implement different defensive strategies in response. For instance, an immediate measure is to deploy honeyfiles on the vulnerable sensitive hosts identified in Table \ref{tab:hosts_info}. After the introduction of these new honeyfiles, the attacking agent's performance significantly deteriorated. Out of 100 episodes, it averaged a reward of -197 and encrypted only 2.6 hosts per episode, compared to the previous averages of 4818 reward and 10.9 encrypted hosts per episode, respectively (refer to Table \ref{tab:sensitivity}).

However, this remedy may offer only temporary protection, assuming the attacker possesses some capabilities \footnote{In the experiment, it is assumed that the agent can identify honeyfiles with a probability of 0.8 after performing a file scanning action.} to detect honeyfiles. 
Upon re-training the RL agent in this altered environment, it partially recovers its previous performance, achieving an average reward close to 3000 and encrypting 5.3 hosts per episode (Table \ref{tab:sensitivity}). 
An example attack path is shown in Table \ref{tab:sample_path}. It is evident that the agent has learned to conduct file-scans before attempting to encrypt sensitive hosts. Despite incurring additional costs, these additional steps represent essential precautions that the agent must take when conducting attacks in a more fortified environment.

Table \ref{tab:sample_path} also highlights the key pivot point in the attack path, namely host (18, 4), which runs multiple services including OpenSSH, HTTP, HTTPS, and MS-SQL-M. Gaining control of this host enables communication to the other subnets, a capability not feasible from the starting DMZ (i.e., subnet 15), which is only connected to subnets 5, 6, 12, 18, 19, and 20. These findings suggest several solutions to enhance network security. For instance, strengthening access control and promptly applying patches to address vulnerabilities at critical junctures, as well as refining network segmentation to better isolate critical assets from less secure areas. Once these improvements are made, RL agents can again be trained to further evaluate their effectiveness. As a result, the iterative process between blue and red teams can facilitate the building of a more robust and secure network environment.

\begin{table}[t]
    \centering
    \caption{Comparison of performance (mean $\pm$ sd) on the original network, modified network with new honeyfiles, and retrained model on the new network ($\rho=1$).}
    \begin{tabular}{c|c|c|c}
    \hline    
     & Original model & New honeyfiles  & Retrained model\\ \hline
     Rewards&  4818 $\pm$ 356 & -197 $\pm$ 563 &  2996 $\pm$ 1126\\
     Steps & 200 $\pm$ 35 & 109 $\pm$ 45 & 172 $\pm$ 61\\
     Compromised & 39.2 $\pm$ 5.8 & 26.0 $\pm$ 7.0 & 29.2 $\pm$ 8.8\\
     Encrypted & 10.9 $\pm$ 1.9 &  2.6 $\pm$ 2.3 & 5.3 $\pm$ 1.4\\
     \hline
    \end{tabular}
    \label{tab:sensitivity}
\end{table}

\begin{table}[t]
    \centering
    \caption{Example attack path.}
    \begin{tabular}{c|c|c} 
    \hline
        Steps & Action  & Target \\ \hline
         1 & Subnet Scan & (15, 12)\\
         2 & Exploit & (18, 4) \\
         3 & Subnet Scan & (18, 4) \\
         4 & Exploit & (12, 0) \\
         5 & File Scan & (12, 0) \\
         6 & Exploit & (10, 5) \\
         7 & Exploit & (9, 0) \\
         8 & File Scan & (9, 0) \\
         9 & File Scan & (10, 5) \\
         10 & Exploit & (8, 0) \\
         11 & File Scan & (8, 0) \\
         12 & Exploit & (21, 3) \\
         13 & File Scan & (15, 12) \\
         14 & File Scan & (21, 3) \\
         15 & Encrypt & (10, 5) \\
         16 & Encrypt & (9, 0) \\
         17 & Exploit & (16, 7) \\
         18 & Encrypt & (8, 0) \\
         19 & Exploit & (15, 14) \\
         20 & Encrypt & (21, 3) \\
         21 & Encrypt & (12, 0)\\ 
         \hline
    \end{tabular}
    \label{tab:sample_path} 
\end{table}

\section{Conclusion} \label{sec:conclusion}
We have proposed a ransomware attack simulation and its RL model as a red teaming tool to streamline the penetration testing process. By leveraging RL, potential attacking strategies or pathways can be quickly identified. We demonstrate this on a medium-sized network, where effective attack strategies with varying risk aversion factors can be learned. The RL agent is capable of discovering high-value targets within the network and coordinating its attacks to maximize damage.
These findings provide valuable insights and can help blue teams develop more resilient defensive measures, which may, in turn, be used to train more advanced attacking agents, thereby revealing additional weaknesses in the security posture.

One extension of the current work is to consider the adversarial multi-agent RL framework, where the attacker and defender compete with each other directly in the same environment. Additionally, instead of statically placing honeyfiles on the same hosts, defenders may opt to deploy honeyfiles or honeypots dynamically to distract or deter attackers.

Another direction for future work is to integrate the attack model with the multi-objective reinforcement learning (MORL) framework, which allows the RL agent to optimize multiple (and potentially conflicting) objectives simultaneously, thus enabling it to handle complex decision-making scenarios. Unlike the current RL formulation, which requires meticulous crafting of rewards and poses hard thresholds on termination conditions, MORL offers a multi-dimensional reward function that can directly incorporate various metrics of the attack campaign. This will also provide greater flexibility to end-users based on their preferences.


\bibliographystyle{IEEEtran}
\bibliography{ref}

\end{document}